\documentclass{article}
\usepackage{spconf,amsmath,epsfig}
\usepackage{amsmath,amssymb,amsfonts}
\usepackage{algorithmic}
\usepackage{graphicx}
\usepackage{tabularx}
\usepackage{multirow}
\usepackage{subcaption}
\usepackage{textcomp}
\usepackage{xcolor}
\usepackage[export]{adjustbox}
\usepackage{float}
\usepackage{makecell}%

\usepackage{color}
\usepackage{siunitx}

\usepackage{adjustbox,booktabs,tabularx}
\usepackage{scalerel}
\usepackage{tikz}
\usepackage{hyperref}

\usetikzlibrary{shapes,positioning}

\usetikzlibrary{svg.path}

\title{Shallow Water Bathymetry Survey using an Autonomous Surface Vehicle}
\name{Bibin Wilson\textsuperscript{*}, Anand Singh\textsuperscript{*}, Amit Sethi\textsuperscript{\textdagger}}
\address{{\textsuperscript{*}Department of Earth Sciences,
	 \textsuperscript{\textdagger}}Department of Electrical Engineering\\Indian Institute of Technology Bombay, Maharashtra, India}

\begin{document}
%
\maketitle
\begin{abstract}
Accurate and cost effective mapping of water bodies has an enormous significance for environmental understanding and navigation. However, the quantity and quality of information we acquire from such environmental features is limited by various factors, including cost, time, security, and the capabilities of existing data collection techniques. Measurement of water depth is an important part of such mapping, particularly in shallow locations that could provide navigational risk or have important ecological functions. Erosion and deposition at these locations, for example, due to storms and erosion, can cause rapid changes that require repeated measurements. In this paper, we describe a low-cost, resilient, unmanned autonomous surface vehicle for bathymetry data collection using side-scan sonar. We discuss the adaptation of equipment and sensors for the collection of navigation, control, and bathymetry data and also give an overview of the vehicle setup. This autonomous surface vehicle has been used to collect bathymetry from the Powai Lake in Mumbai, India.

\end{abstract}
\section{Introduction and Background}
\label{sec:intro}
At least since 1800 BC, attempts have been made to measure the seabed. While currently, bathymetry is employed primarily to measure ocean depth, many other use cases exist for lakes, dams, rivers, and other freshwater basins  \cite{hare2008small}. Mapping of hydroelectric power plants, whose infrastructures must be inspected regularly, is a typical example of such a use case. The total volume and depth distribution, particularly around the dam discharge and the submerged spillway equipment, are of significant importance \cite{bourgeois1999autonomous}. Monitoring changes in ecologically sensitive water bodies that are under anthropogenic threat is another such use case. While considerable research has been done for deep water bodies, not much attention has been paid to efficient mapping of shallow water bodies using cost-effective means.

Remote sensing is the acquisition, without physical touch, of information about a phenomenon or an object. Remote sensing is used in several areas, such as military intelligence, planning, geographical surveys, and ecological understanding \cite{campbell2011introduction}. Remote Remote sensing helps to collect rapid and cost-effective data in large areas without disrupting ecology or geology. The phrase "remote sensing" refers primarily to satellite or aerial sensing technology used to detect or classify earthly objects based on electromagnetic signals. Orbital platforms gather and transmit electromagnetic spectrum data that provide researchers with adequate data to follow patterns such as catastrophes or natural calamities and other occurrences. But, in terms of space, spectrum, and radiometry, the major problem of such approaches is the poor resolution of the data. On the other hand, traditional manual survey is time-consuming, labor-intensive, and costly. We thus require new ways or tools to measure water bodies carefully \cite{hudson2014underway}. Unmanned and autonomous vehicles can help solve these problems and allow scientists and researchers to understand and minimize the consequences of a constantly changing environment.

To be autonomous, a vehicle must function without external help or control to navigate and perform its intended operations. For the current prototypes of autonomous aquatic vehicles, there are numerous parameters and the design varies considerably depending on the applications. A typical unit has a hull, a propulsion system, a navigation system, and a data collection and transmission system \cite{bourgeois1999autonomous}. An ASV (autonomous surface vehicle) differs from a USV (unmanned surface vehicle) in that the former does not have a remote driver who operates the vehicle. ASVs can perform its function, such as navigation and data collection without a remote pilot, and can either transmit the data to a home base or store the data on board.

For side-scan sonar surveys, the usage of a non-ferrous vessel and installation of the sensors linked to the ship's hull are more practical for shallow-water bodies. Hare et al. \cite{hare2008small} illustrates different adjustments to shallow water applications of traditional mid-size and small survey boats. However, complicated ship designs have limited utility in shallow waters and restricted places near the shores. The use of surface vehicles, as opposed to a hydrographic vessel capable of taking measurements in shallow seas, is an alternate method, which is being employed more because of continuous technological advancements. Specht et al. \cite{specht2017application} describe the idea of bathymetric measurements for shallow seas using an independent, unmanned survey vessel. Due to their relevance for the safety of navigation and transit, the focus will be on developing bathymetric charts, especially in the coastal area.

Suhari et al. \cite{suhari2017small} outline the transition from the bathymetric survey to a bathymetric vehicle utilized for the examination of the seafloor topography utilizing hydrographic measures. The concentration is on inland water and lake surveys. Similarly, the gathering of data on bathymetry and environmental variables in shallow seas through the use of several sensors in tiny surface vehicles is demonstrated by Giordano et al. \cite{giordano2016integrating}. This article provides several advances for guiding, navigating and controlling USV so that autonomous survey campaigns can be carried out in shallow seas. 

We developed a cost-effective and autonomous solution for mapping shallow-water bodies using a side-scan sonar. We describe its design and the data collection procedure.

\section{Materials and Method} \label{sec:format}

\subsection{Design of the vehicle}
\begin{figure}
  \centering
  \begin{subfigure}[t]{.45\linewidth}
    \centering\includegraphics[width=.99\linewidth]{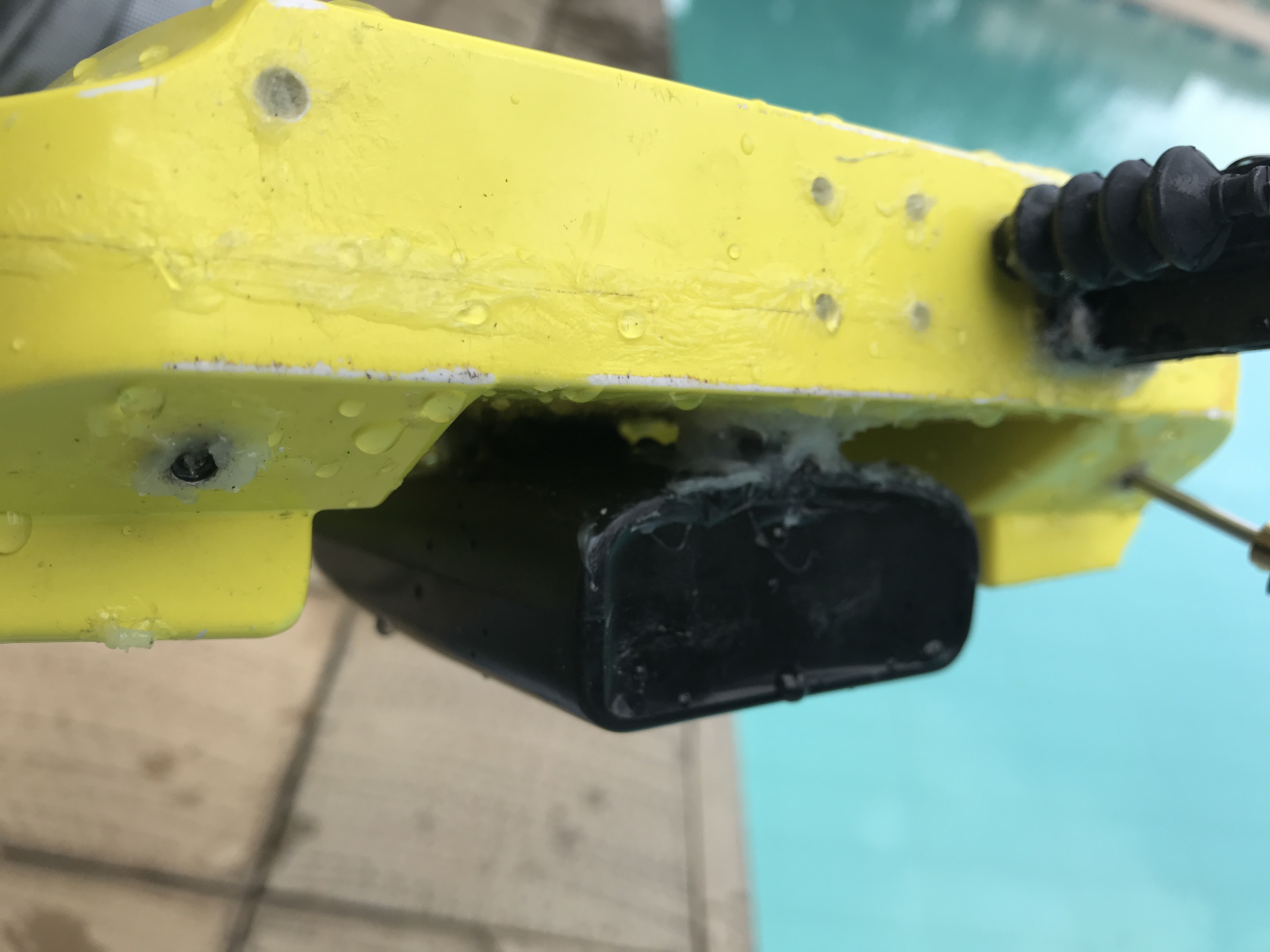}
    \caption{Sonar transducer fitted in hull}
    \label{fig_datasets:sub1}
  \end{subfigure}
  \medskip
  \begin{subfigure}[t]{.45\linewidth}
    \centering\includegraphics[width=.99\linewidth]{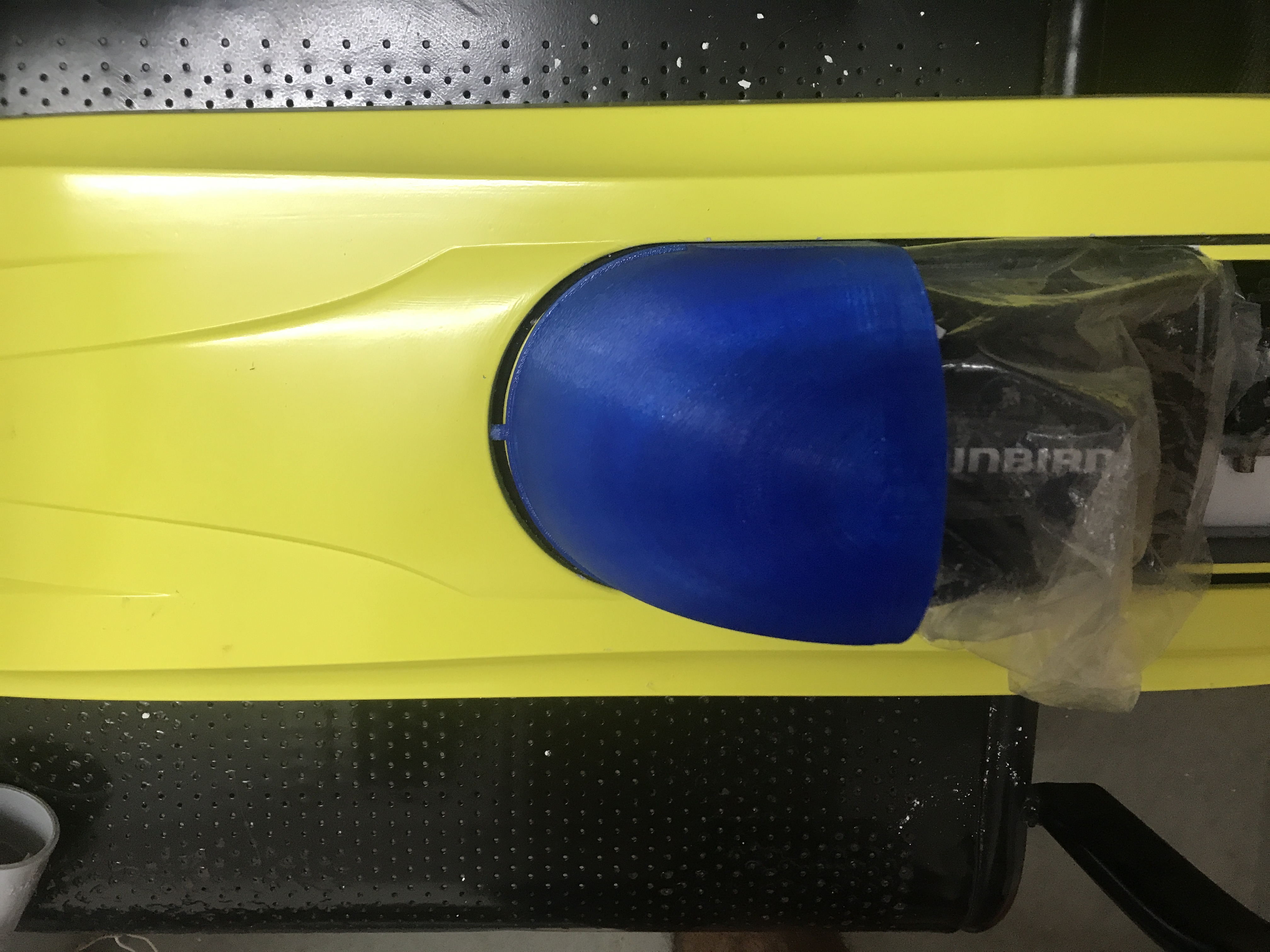}
    \caption{Sonar head in hull}
    \label{fig_datasets:sub2}
  \end{subfigure}
  \begin{subfigure}[t]{.45\linewidth}
    \centering\includegraphics[width=.99\linewidth]{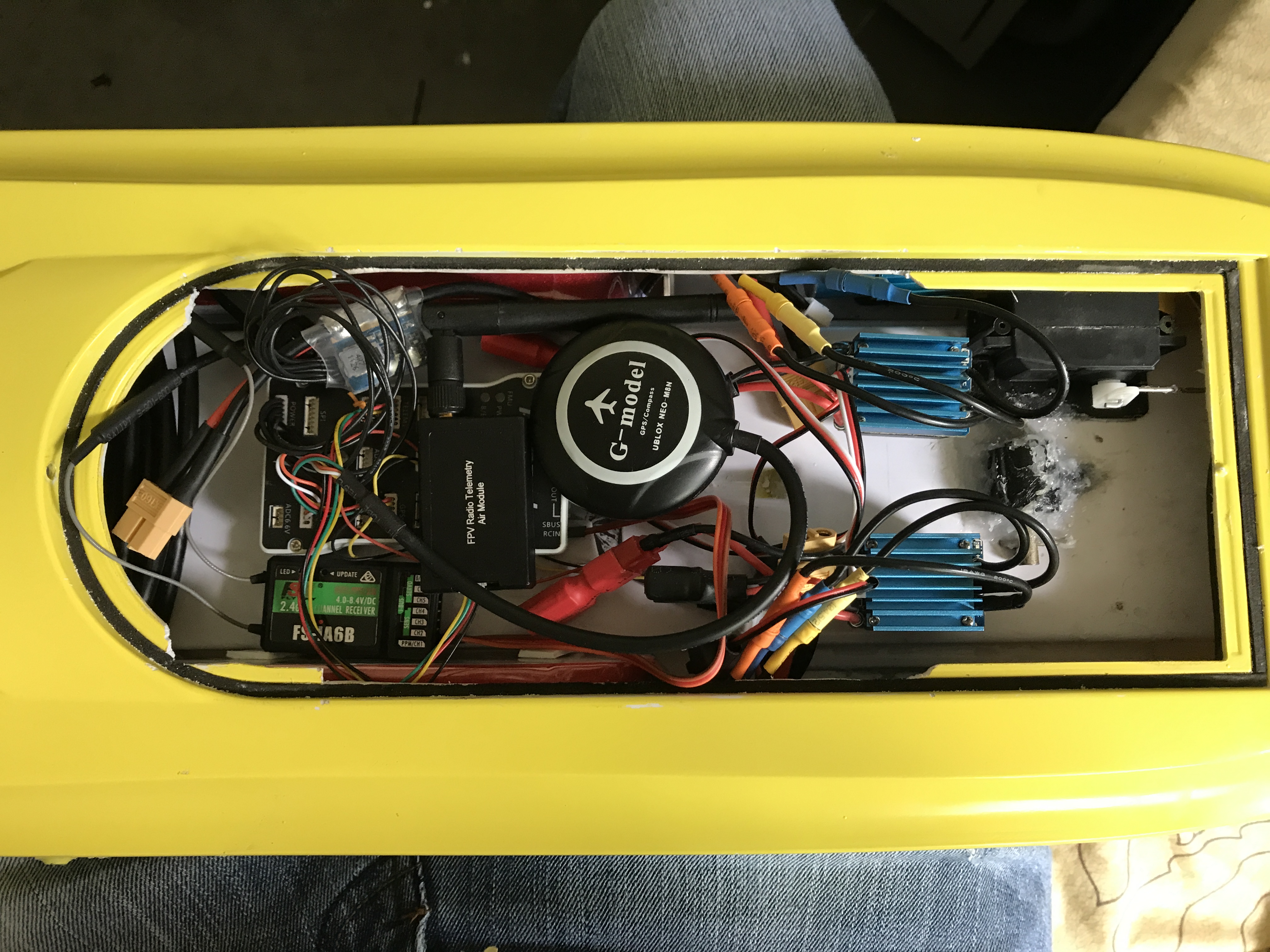}
    \caption{Electronics in testing stage}
    \label{fig_datasets:sub3}
  \end{subfigure}
  \caption{ASV in testing stages}
  \label{fig_datasets}
  
\end{figure}

\begin{figure}
  \centering
  \begin{subfigure}[t]{.45\linewidth}
    \centering\includegraphics[height=3cm]{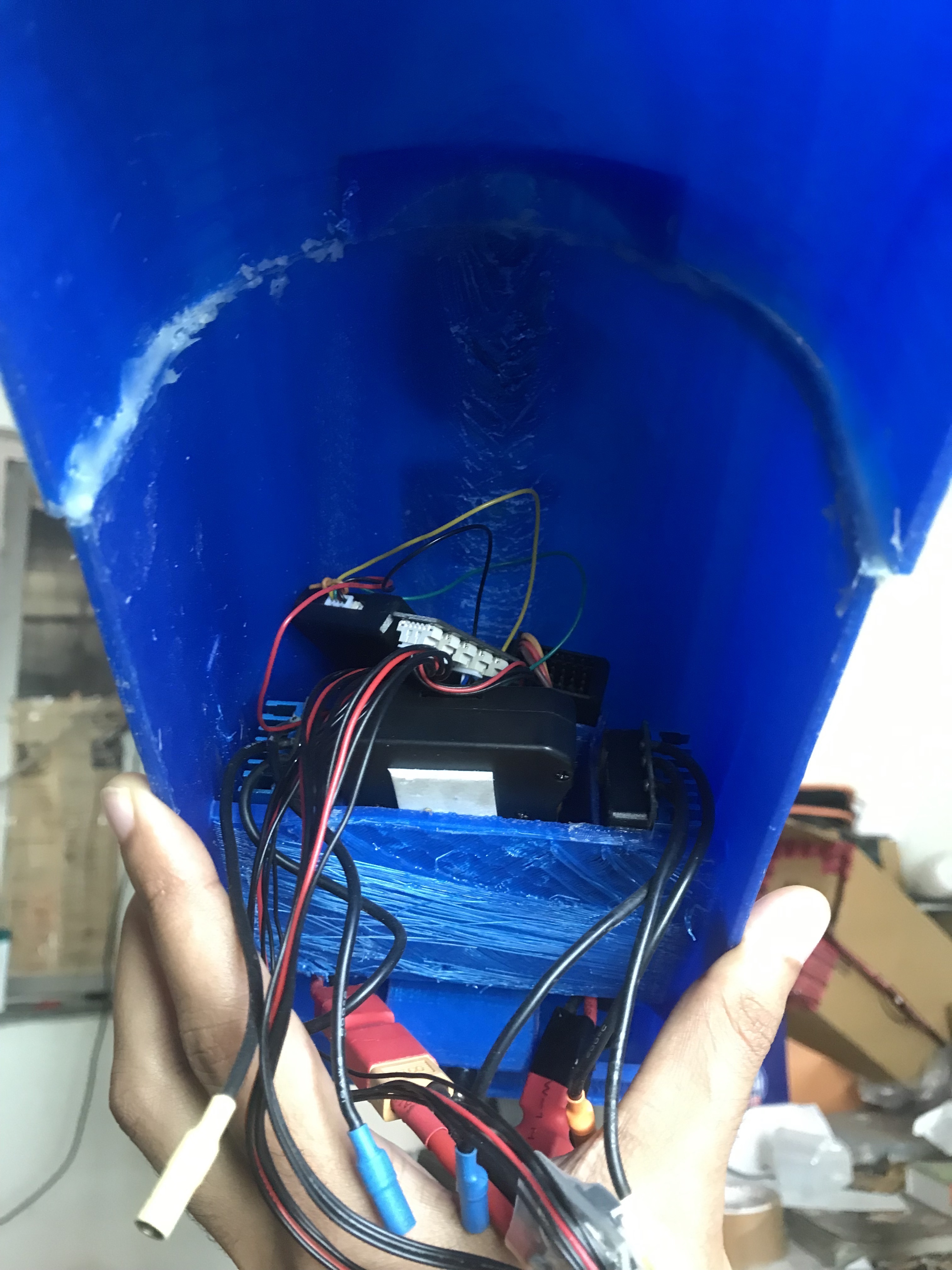}
    \caption{Inside View}
    \label{fig_datasets:sub3}
  \end{subfigure}
  \medskip
  \begin{subfigure}[t]{.45\linewidth}
    \centering\includegraphics[height=3cm]{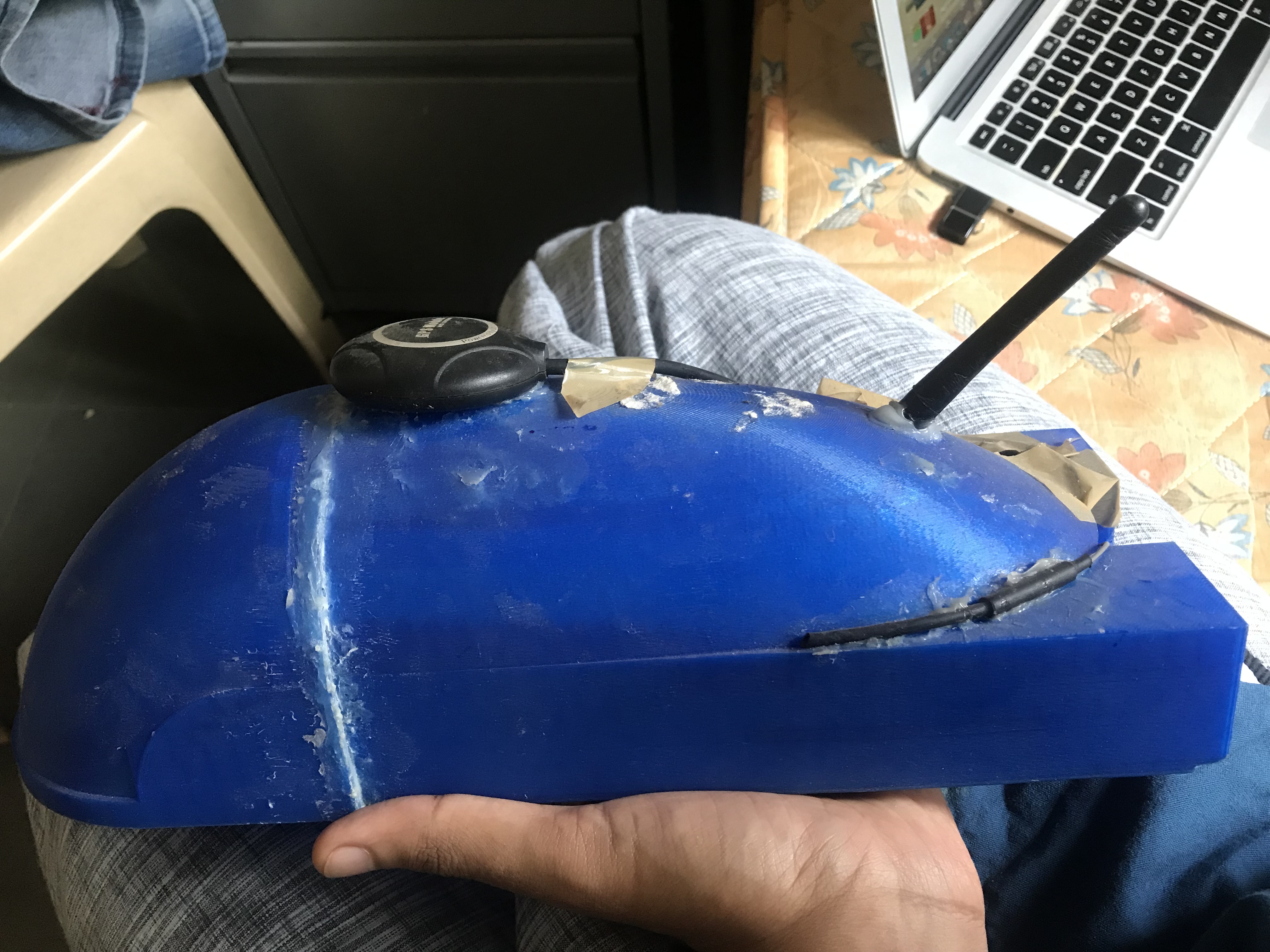}
    \caption{Outside View}
    \label{fig_datasets:sub4}
  \end{subfigure}
  \caption{3D printed part with component placing }
  \label{fig_datasets2}
  
\end{figure}

Catamaran structures have higher stability due to the wider beams, lesser power requirements due to smaller hydrodynamic resistance, and shallower draught, which helps in shallower areas. \cite{stanghellini2020openswap}  Since the acoustic and ultrasonic signal generated by the sonar transducer should be perfectly coupled with the water and the sensor should also be shielded from the effect of turbulence, the design of the components, propellants, and sensor placements was planned accordingly as shown in Figures~\ref{fig_datasets} and~\ref{fig_datasets2}. 

To ease the implementation, we selected a readily available catamaran hull, and 3D printed the required structure on them to safely incorporate the electronic components. 

\subsection{Electronic Components}
Main electronics components used for this study in Table~\ref{tab:Electronic-Component}

\begin{table}[]
\caption{Main Electronic Components}
\label{tab:Electronic-Component}
\begin{tabular}{ll}
\toprule
\textbf{Component}               & \textbf{Specification}                \\
\midrule
Sonar                   & CHIRP SI GPS G2              \\
Controller              & PixHawk PX4                  \\
Positioning System      & Ublox Neo-M8N GNSS           \\
Motor with ESC          & 3100/3930KV Brushless        \\
Power Supply            & 8000mAh Lipo \& 18650 Li-ion \\
RC Transmitter Receiver & 2.4GHz 6CH AFHDS             \\
\bottomrule
\end{tabular}
\end{table}

\subsubsection{Humminbird Side Imaging (SI) Sonar}
This study used the Humminbird Helix 5 CHIRP SI GPS G2 to record the bathymetry data. The instrument comes with a transducer and a control head that have three options for underwater imaging: down imaging, dual beam, and side imaging. It is capable of 3 frequencies: 83, 200, and 455 kHz. 

For dual beam, the transducer projects two concentric conical sonar beams onto water directly under the transducer at an angle of $20^{\deg}$ (200 kHz) and $60^{\deg}$ (83 kHz) openings. The down-imagery uses a razor-thin, high-frequency down-looking sonar beam to create a 2D profile of the waterbed. For side imaging, the sonar beams are thin-shaped, with angles of $86^0$, and could reach a range of 73 meters side to side.

CHIRP technology allows the sonar pulse to be modulated in a range of frequencies. For 2D profiling, frequency ranges of 75-95 kHz and 175-225 kHz were used. For down imaging and side imaging, modulation of 420-520 kHz was used. The unit has a maximum depth capability of 457 m. The unit is equipped with an internal precision GPS module for geolocation tagging. 

\subsubsection{Controller:PixHawk}

PixHawk internal software architecture consists of two main layers: flight stack and middleware. The flight stack is an autonomous control and estimation system and the middleware is the general robotic layer that can support any autonomous robot providing internal-external communication and hardware integration.

\begin{figure}[htb]

\begin{minipage}[b]{1.0\linewidth}
  \centering
  \centering\includegraphics[width=.96\linewidth]{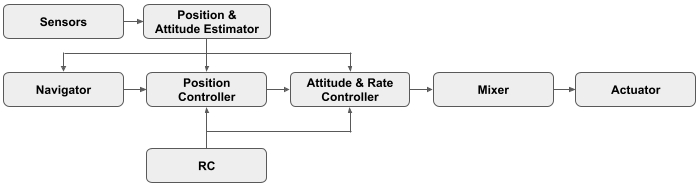}
\end{minipage}
\caption{overview of the flight stack.}
\label{fig:flight-stack}
\end{figure}

An overview of the flight stack is shown in Figure~\ref{fig:flight-stack}. The estimator combines one or more sensor inputs (such as GPS, IMU, etc.) and computes the vehicle state. The position controller takes the inputs from the sensors (through the estimator), the navigator (such as autonomous flight controller), and the remote controller (RC) inputs and takes the decision and drives that to actuators, such as motors or servo controllers. A controller’s goal is to adjust the values of measurement(estimated state) such that it matches the position setpoints (desired position) from the navigator or RC. The output from the controller is a correction to eventually reach that set point. The mixer takes the commands (such as turning right or left, etc.) from the controller and translates them into the individual motors. Tuning the variables of the controller such as P, I, D, maximum throttle, angle, cruise speed, acceleration, etc. ensures that the vehicle works properly in real-world conditions.

\subsubsection{Positioning system}
To provide the position during the survey to the ASV, Ublox Neo-M8N, a separate GNSS module with a digital compass (HMC5883L) used in conjunction with PixHauk. This GNSS module offers the capability to track the satellite constellations including GPS, GLONASS, Galileo, BeiDou, QZSS, and SBAS.


\subsection{Sensor calibration}
We calibrated the sonar and controller in the controlled environment of a swimming pool. The depth error was corrected using the known depth of the swimming pool and was given as an offset as shown below: 
\begin{equation}
    error = Depth_{sonar} - Depth_{known}
\end{equation}

Using Mission Planner software, the parameters in steering rate, steering modes, speed, throttle, motors, and navigation were tuned. This includes P, I, D values, turn radius, max acceleration, speed, etc.

\begin{figure}
  \centering
  \begin{subfigure}[t]{.49\linewidth}
    \centering\includegraphics[height=3cm]{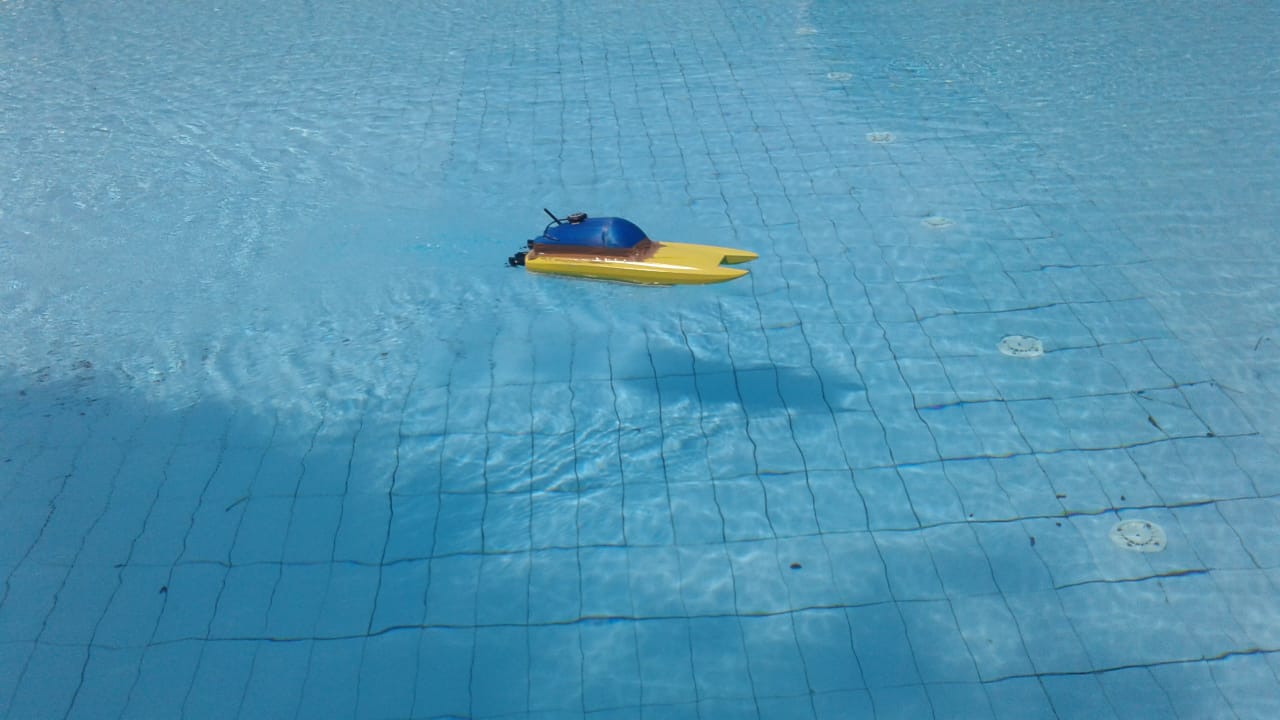}
    \caption{ASV in Swimming Pool}
    \label{fig_datasets:sub1}
  \end{subfigure}
  \medskip
  \begin{subfigure}[t]{.49\linewidth}
    \centering\includegraphics[height=3cm]{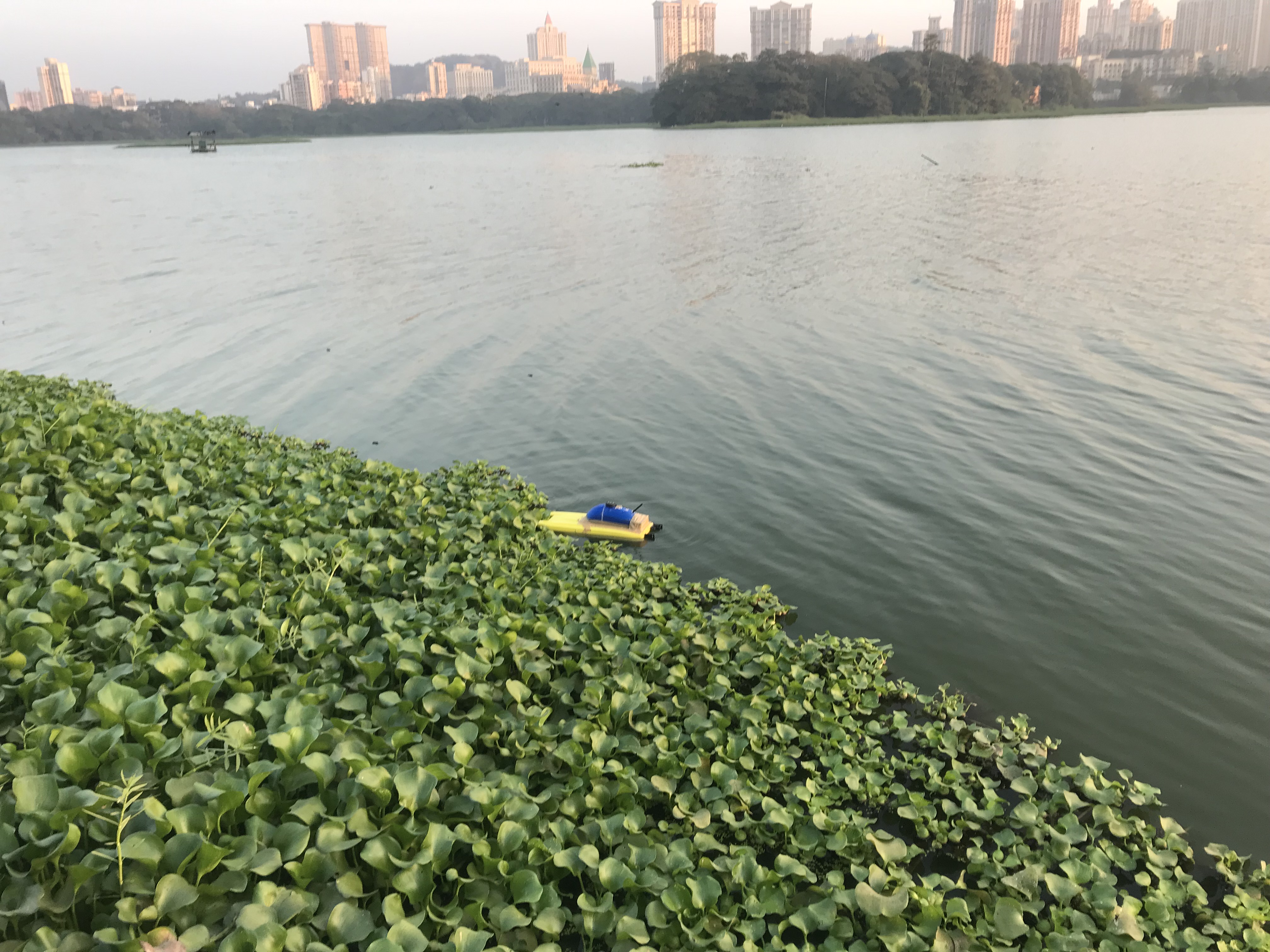}
    \caption{ASV in Powai Lake}
    \label{fig_datasets:sub2}
  \end{subfigure}
  \caption{ASV in data collection}
  \label{fig_datasets3}
  
\end{figure}

\section{Data Acquisition}
\subsection{Study Area}
The study area was Powai Lake located in Mumbai, India, which is an artificial freshwater lake. The lake spreads over around 2 square kilometers from N019.08.140, E072.53.740 in North-West to N019.07.176, E072.54.829 South-East.


Data acquisition was done during summer, in the month of March. The presence of aquatic weeds such as water hyacinth and water lettuce at the time of the survey limited us from covering the lake in whole. 

\begin{figure} 
  \centering
  \begin{subfigure}[t]{.49\linewidth}
    \centering\includegraphics[height=3cm]{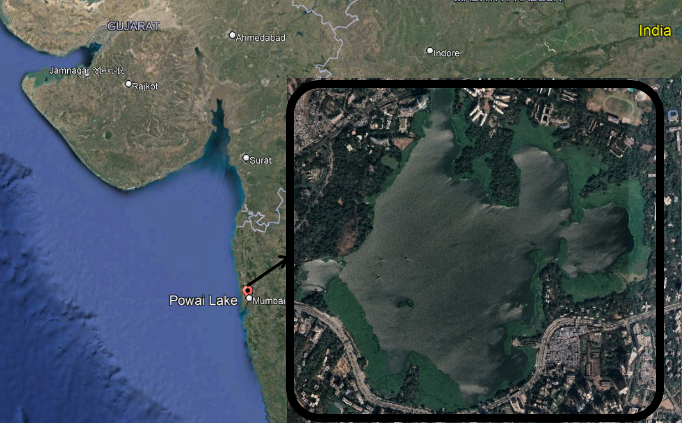}
    \caption{Powai Lake}
    \label{fig_datasets:sub1}
  \end{subfigure}
  \medskip
  \begin{subfigure}[t]{.49\linewidth}
    \centering\includegraphics[height=3cm]{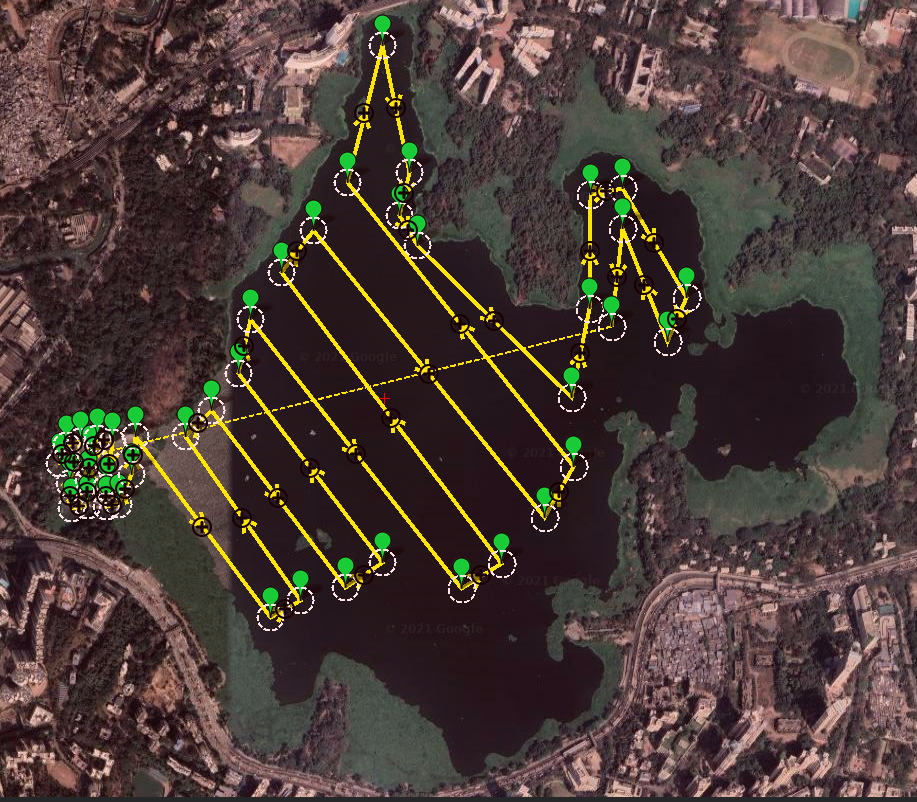}
    \caption{Path Planned}
    \label{fig_datasets:sub2}
  \end{subfigure}
  \caption{Powai Lake: Study Area}
  \label{fig:powai-depth-map}
  
\end{figure}

\section{Results and Discussions}
Obtained sonar files in the format of *.SON, *.IDX, and *.DAT were viewed and interpreted using the software “ReefMaster”. A final map is obtained as shown in Figure~\ref{fig:powai-depth-map} after interpolating the data points using TIN (triangulated irregular network). Here, the data points were triangulated and the values between the data points were interpolated using the slope of the connecting triangles.

\begin{figure}[htb]

\begin{minipage}[b]{1.0\linewidth}
  \centering
  \centering\includegraphics[width=.96\linewidth]{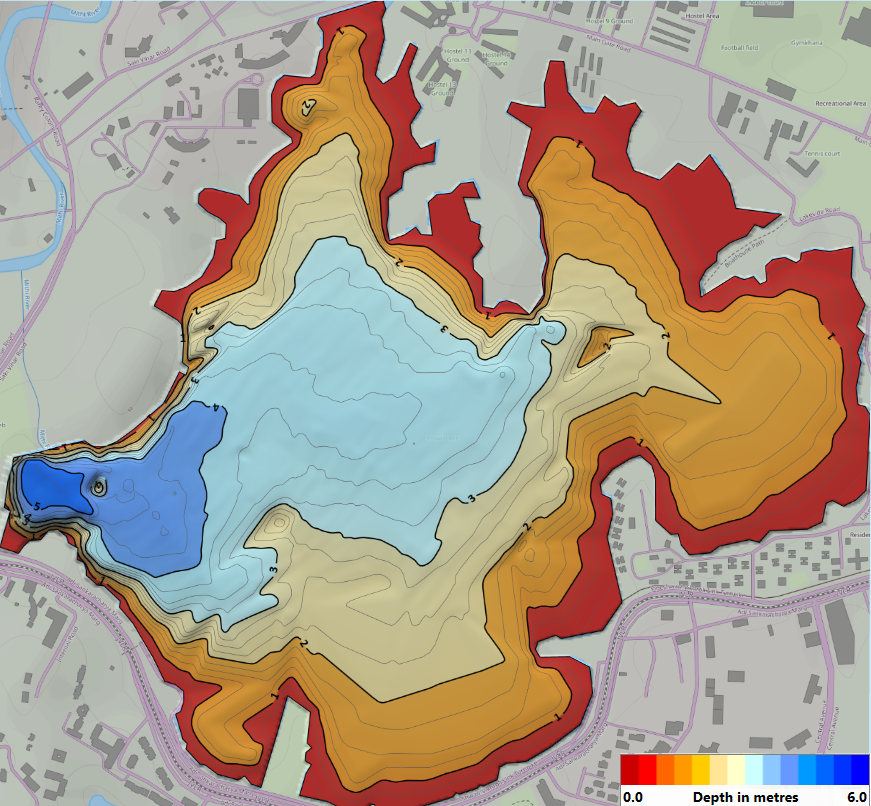}
\end{minipage}
\caption{Depth Map Generated}
\label{fig:powai-depth-map}
\end{figure}

  


\begin{table}[]
\caption{Water Volume}
\label{tab:waterVolume}
\begin{tabular}{llll}
\toprule
\textbf{Lower} (m) & \textbf{Upper} (m) & \textbf{Volume ($m^3$)} & \textbf{Area ($m^2$)} \\
\midrule
0.00      & 1.00      & 1651393.27     & 1765845.00   \\
1.00      & 2.00      & 1143354.21     & 1407421.00   \\
2.00      & 3.00      & 657685.43      & 857950.00    \\
3.00      & 4.00      & 290903.70      & 485206.00    \\
4.00      & 5.00      & 37717.81       & 91768.00     \\
5.00      & 6.00      & 1911.85        & 12448.00    \\
\bottomrule
\end{tabular}
\end{table}

Table \ref{tab:waterVolume} shows the volume and surface area covered by each depth area in a 1-meter interval. For example, an area of 12448 $m^2$ is present within the depth range of 5 to 6 meters, and that area has a volume capacity of 1911.85 $m^3$.

The following findings were obtained from the survey of the Powai lake, after the interpolated depth:
\begin{itemize}

    \item Total Volume: 3782966 $m^3$
    \item Total Mapped Area: 1765845 $m^2$
    \item Average Depth : 2.1m
    \item Maximum Depth : 5.83m
\end{itemize}

\section{Conclusion and Future Work}
\label{sec:typestyle}

The results of the survey show that this technique is suitable for monitoring in shallow areas. These shallow areas, especially during storms, are rapidly changing due to erosion and deposition, as well as human activity. Because of the difficulty in navigating in low water and the large space sample rates necessary, such water bodies present a tough problem for standard ship-sounding.This catamaran-based autonomous surface vehicle can assist in investigating the implications of a changing environment for such shallow water bodies.


The independent system can remain on duty as long as it has enough energy. The autonomous system can adjust on site to environmental changes itself or to survey tasks with advanced mission planning and decision-making capabilities on board. The modest size of this model decreases the likelihood of wind catches, since the top half is aerodynamic and has little surface presence. The stability and shape of the streamline also assist the model to cope with difficult environmental fluctuations and also enhance autonomous surface vehicle management. 

This research paper introduced a low-cost robust autonomous surface vehicle for bathymetry collection. It should also be noted that although such systems can collect data independently, there is a requirement for a considerable amount of human labor and resources to manage and implement such systems. New sensors and technologies can be integrated into the system to improve vehicle independence and efficiency.


%
%
%


\bibliographystyle{IEEEbib}
\bibliography{strings,refs}

\end{document}